%Paper: hep-ph/9303280
%From: DIAZMA@ctrvax.Vanderbilt.Edu
%Date: 18 Mar 1993 16:35:57 -0600 (CST)

%%macropackage=phyzzx
\input phyzzx.tex
\input myphyx.tex
\overfullrule0pt

%\physrev

\def\etal{{\it et al.}}

\def\bold#1{\setbox0=\hbox{$#1$}%
     \kern-.025em\copy0\kern-\wd0
     \kern.05em\copy0\kern-\wd0
     \kern-.025em\raise.0433em\box0 }
\Pubnum={VAND-TH-93-2}
\date={January 1993}
\pubtype{}
\titlepage

% para publicacion:
%\twelvepoint
%\doublespace
%\vsize=   8.00in
%\hsize=   6.00in
%\hoffset=  .25in
%\voffset= 0.25in

\vskip1cm
\title{\bf Constraints on Supersymmetry due to
$b\longrightarrow s\gamma$\ : an Improved Calculation}
\author{Marco Aurelio D\'\i az }
\vskip .1in
\centerline{Department of Physics and Astronomy}
\centerline{Vanderbilt University, Nashville, TN 37235}
\vskip .2in

\centerline{\bf Abstract}
\vskip .1in

The calculation of the one-loop induced decay rate $b\longrightarrow
s\gamma$ is improved by including radiative corrections to the
charged Higgs mass and to the charged-Higgs-fermion-fermion vertex.
It is shown how the radiative corrections modify the excluded zone
in the $\tan\beta-m_A$ plane, imposed by the non-observation of this
decay. In particular, a light CP-odd Higgs at large
$\tan\beta$ becomes allowed.

\vfill

\endpage

%para preprint:
\voffset=-0.2cm

\REF\guide{J.F. Gunion, H.E. Haber, G. Kane and S. Dawson,
{\it The Higgs Hunter's Guide} (Addison-Wesley, Reading MA, 1990).}
\REF\diazhaberi{M.A. D\'\i az and H.E. Haber, {\it Phys. Rev. D}
{\bf 45}, 4246 (1992).}
\REF\CPRbrignole{P.H. Chankowski, S. Pokorski and J. Rosiek,
{\it Phys. Lett. B} {\bf 274}, 191 (1992); A. Brignole, {\it Phys. Lett.
B} {\bf 277}, 313 (1992).}
\REF\diaz{M.A. D\'\i az, Report No. VAND-TH-93-1, January 1993,
unpublished.}
The Minimal Supersymmetric Model (MSSM) is a well motivated
extension of the Standard Model (SM) that has in the Higgs sector
a pair of charged Higgs bosons $H^{\pm}$\refmark\guide. Their tree
level mass is fixed by the CP-odd Higgs mass and by the $W$-boson mass
through the relation: $m_{H^{\pm}}^2=m_A^2+m_W^2$. Nevertheless, it has
been shown that radiative corrections to the charged Higgs mass
can be large if there is a substantial mixing in the squark mass
matrix\refmark{\diazhaberi,\CPRbrignole}. Recently, the radiative
corrections
to the charged-Higgs-fermion-fermion vertex have been calculated
\refmark\diaz. They are also large in the presence of mixing
in the squark mass matrix.

\REF\expbsf{M. Battle \etal, CLEO Collaboration, to appear in the
{\it Proceedings of the joint Lepton-Photon and Europhysics
International Conference on High Energy Physics}, Geneva, Switzerland,
August 1991.}
\REF\hewettBBP{J.L. Hewett, Report No. ANL-HEP-PR-92-110, November 1992,
unpublished; V. Barger, M.S. Berger and R.J.N. Phillips, Report No.
MAD/PH/730, November 1992, unpublished.}
\REF\InamiL{T. Inami and C.S. Lim, {\it Prog. Theor. Phys.} {\bf 65},
297 (1981).}
\REF\TwoHDM{T.G. Rizzo, {\it Phys. Rev. D} {\bf 38}, 820 (1988); B.
Grinstein and M.B. Wise, {\it Phys. Lett. B} {\bf 201}, 274 (1988);
W.-S. Hou and R.S. Willey, {\it Phys. Lett. B} {\bf 202}, 591 (1988);
T.D. Nguyen and G.C. Joshi, {\it Phys. Rev. D} {\bf 37}, 3220 (1988);
C.Q. Geng and J.N. Ng, {\it Phys. Rev. D} {\bf 38}, 2857 (1988);
D. Ciuchini, {\it Mod. Phys. Lett. A} {\bf 4}, 1945 (1989);
B. Grinstein, R. Springer and M. Wise, {\it Nucl. Phys.} {\bf B339},
269 (1990); V. Barger, J.L. Hewett and R.J.N. Phillips, {\it Phys.
Rev. D} {\bf 41}, 3421 (1990), and Erratum.}
\REF\BBMR{S. Bertolini, F. Borzumati, A. Masiero and G. Ridolfi,
{\it Nucl. Phys.} {\bf B353}, 591 (1991).}
\REF\susy{S. Bertolini, F. Borzumati and A. Masiero, {\it Nucl. Phys.}
{\bf B294}, 321 (1987); F.M. Borzumati, Report No. PRINT-93-0025
(Hamburg), unpublished.}
On the other hand, it has been pointed out that the upper
bound on the branching fraction of the inclusive decay $b\rightarrow
s\gamma$, given by $B(b\rightarrow s\gamma)<8.4\times 10^{-4}$
\refmark\expbsf, sets powerful constraints on the charged Higgs sector
\refmark\hewettBBP. This decay is forbidden at tree level,
but induced in the Standard Model (SM) at the one-loop level with W
and Goldstone bosons circulating in the loop\refmark\InamiL. In two
Higgs doublets extensions of the SM, the charged Higgs also contributes
to this decay\refmark\TwoHDM, and in the MSSM, charginos, neutralinos,
gluinos and scalar quarks also have to be included\refmark{\BBMR,\susy}.

\REF\qcd{S. Bertolini, F. Borzumati and A. Masiero, {\it Phys. Rev.
Lett.} {\bf 59}, 180 (1987); N.G. Deshpande, P. Lo, J. Trampetic,
G. Eilam and P. Singer, {\it Phys. Rev. Lett.} {\bf 59}, 183 (1987);
B. Grinstein, R. Springer and M.B. Wise, {\it Phys. Lett. B}
{\bf 202}, 138 (1988); P. Cho and B. Grinstein, {\it Nucl. Phys.}
{\bf B365}, 279 (1991).}
Two-loop corrections to a process that is forbidden at tree level are
potentially important. For example, QCD corrections
to $b\rightarrow s\gamma$ have been
found to be large for a top quark mass not far from its experimental
lower bound\refmark\qcd. Nevertheless, electroweak corrections have
been considered only partially in the charged Higgs mass
\refmark\hewettBBP. Here, we estimate the influence of the electroweak
corrections to the charged Higgs mass and to the
charged-Higgs-fermion-fermion vertex on the decay rate of the process
$b\rightarrow s\gamma$, in the context of the MSSM.

\REF\MPomarol{A. Mendez and A. Pomarol, {\it Phys. Lett. B}
{\bf 265}, 177 (1991).}
The renormalized $H^+d\bar u$ vertex is:
$$\Gamma^{H^+d\bar u}(p^2)=i\lambda^-_{H^+d\bar u}(1-\gamma_5)f^-(p^2)
+i\lambda^+_{H^+d\bar u}(1+\gamma_5)f^+(p^2),\eqn\ecux$$
and the $\lambda^{\pm}$ couplings are defined by:
$$\lambda_{H^+d\bar u}^-={{gm_u}\over{2\sqrt{2}m_Wt_{\beta}}},
\qquad \lambda_{H^+d\bar u}^+={{gm_dt_{\beta}}\over{2\sqrt{2}
m_W}},\eqn\ecuvii$$
Expressions for the factors $f^{\pm}$ are given in ref.~[\diaz]\ valid
for a light pair of fermions ($ud, cs, \nu_e e^-,\nu_{\mu}\mu^-,
\nu_{\tau}\tau^-$). Using these corrections for the $H^+t\bar b$
vertex implies that we are neglecting some terms that, in the context
of a non-supersymmetric two Higgs doublet model, are reported to be
small\refmark\MPomarol.

Following ref.~[\BBMR], at the leading order, the SM and the charged
Higgs contributions to the amplitude of the process $b\rightarrow
s\gamma$ are proportional to the operator $m_b\epsilon_{\mu}\bar s
i\sigma^{\mu\nu}q_{\nu}P_Rb$, where $\sigma^{\mu\nu}=(i/2)[\gamma
^{\mu},\gamma^{\nu}]$ and $q$ and $\epsilon$ are
the four-momentum and polarization of the outgoing
photon. The coefficient $A_{tot}$ is, omitting the contribution of the
supersymmetric partners, given by the sum of eqs. 40\ and 41 in
ref.~[\BBMR]: $A_{tot}=A_{SM}+A_{H^{\pm}}$. The charged Higgs coefficient
can be decomposed into two pieces. One is independent of $\tan\beta$
and the other is proportional to $\cot^2\beta$. Introducing
appropriately the renormalization factors $f^{\pm}$, the corrected
amplitude looks like:
$$A_{tot}=A_{SM}+(f^+f^-)A_{H^{\pm}}^1+(f^-)^2\cot^2\beta
A_{H^{\pm}}^2\eqn\corratot$$
where we evaluate the functions $f^{\pm}(p^2)$ at the typical energy
scale of this decay process $p^2=m_b^2$. Nevertheless,
it was checked that over a range from $m_b$ to several hundreds of
GeV, the dependence
of $f^{\pm}$ on $p^2$ is so weak that the answer is insensitive to
the choice of $p^2$. This brings the calculation closer to what
would be the exact two-loop result.
QCD corrections are also implemented following
the prescription in ref.~[\BBMR].

The branching ratio calculated in this way will depend strongly on
$\tan\beta$, $m_A$ (through the charged Higgs mass) and $m_t$. For
some values of these parameters, the branching ratio is larger than
the experimental upper bound and this effect is represented by an
excluded zone in the plane $\tan\beta-m_A$. For a value of the top
quark mass given by $m_t=150$ GeV, we plot in Fig. 1 contours of
constant value of the branching ratio $B(b\rightarrow s\gamma)$.
In order to fully appreciate the effect of the partial two-loop
electroweak radiative corrections implemented in this calculation,
we plot the pure one-loop answer (dot-dashed line), one-loop plus
corrections only to the charged Higgs mass (dashed line), and
corrections to both the charged Higgs mass and the
charged-Higgs-fermion-fermion
vertex (solid line). Three sets of curves are
presented: one set for $B(b\rightarrow s\gamma)=8.4\times10^{-4}$
(the experimental upper bound), and two other sets for
$B=7.0\times10^{-4}$ and $B=5.5\times10^{-4}$, to illustrate
the dramatc effect of a future measurement of this branching ratio,
or an improved upper bound.

The excluded zone lies below the solid line corresponding to
$B=8.4\times10^{-4}$. We see that the inclusion
of electroweak radiative corrections to the Higgs sector has an
important effect when there is a substantial mixing in the squark
mass matrix. In particular, we cannot rule out a light CP-odd Higgs
since for $\tan\beta\gsim 24$ (this value will change for other
choices of $m_t$ and the squark mixing), the experimental
upper bound of $B(b\rightarrow s\gamma)$ is consistent with any
value of $m_A$. Despite this effect, the decay $b\rightarrow s\gamma$
imposes severe restrictions to the MSSM, and a future measurement of
its branching ratio, or a new upper limit, will rule out a bigger
region in the parameter space, as we can see from the two upper sets
of curves in Fig. 1. The issue of
how these constraints are affected by including the supersymmetric
particles neglected in this calculation remains to be addressed.

{\bf \noindent Acknowledgements }
\vglue 0.4cm
Discussions with Howard Baer and Thomas Weiler are
gratefully acknowledged. This work was supported in part by the U.S.
Department of Energy.

\refout

\FIG\funo{Contours of constant branching ratio $B(b\rightarrow s\gamma)$
in the $\tan\beta-m_A$ plane. We display three values for the
branching ratio:
$8.4\times10^{-4}$, $7.0\times10^{-4}$ and $5.5\times10^{-4}$.
We plot the branching ratio including corrections to the charged
Higgs mass and to the charged-Higgs-fermion-fermion vertex
(solid line),
only corrections to the charged Higgs mass (dashed line),
and the pure one-loop answer with tree level mass and coupling
(dot-dashed line).
All the supersymmetry breaking mass parameters in the squark sector
are taken to be equal to $M_{SUSY}$.}

\figout

\end